\documentclass[letterpaper]{article} 
\usepackage[preprint]{aaai2027}
\usepackage[hyphens]{url}  
\usepackage{graphicx} 
\usepackage{natbib}  
\usepackage{caption} 
\usepackage{algorithm}
\usepackage{algorithmic}

\usepackage{newfloat}
\usepackage{tabularx}
\usepackage{amssymb}
\usepackage{listings}
\DeclareCaptionStyle{ruled}{labelfont=normalfont,labelsep=colon,strut=off} 
\floatstyle{ruled}
\newfloat{listing}{tb}{lst}{}
\floatname{listing}{Listing}

\usepackage{booktabs}
\usepackage{xcolor}

\definecolor{quotebg}{HTML}{F8F7F4}

\newcommand{\mypara}[1]{%
  \noindent
  \colorbox{quotebg}{%
    \strut\textbf{\textit{#1}}%
  }%
}

\usepackage[most]{tcolorbox}

\newcommand{\segabench}{\textsc{SeGaBench}}

\definecolor{quotegold}{HTML}{B08D57}
\definecolor{quotebg}{HTML}{F8F7F4}
\definecolor{quotetext}{HTML}{303030}

\newtcolorbox{highquote}[1][]{%
  enhanced,
  breakable,
  colback=quotebg,
  colframe=quotebg,
  coltext=quotetext,
  boxrule=0pt,
  borderline west={2.5pt}{0pt}{quotegold},
  sharp corners,
  left=4pt,
  right=2pt,
  top=2pt,
  bottom=0pt,
  fontupper=\itshape,
  overlay={
    \node[
      anchor=north west,
      text=quotegold,
      font=\fontsize{28}{28}\selectfont
    ] at ([xshift=4pt,yshift=4pt]frame.north west) {``};
  },
  #1
}

\usepackage{enumitem}

\setlist[itemize]{
  leftmargin=1em,
  label=\raisebox{0.2ex}{\scriptsize$\bullet$},
  itemsep=2pt,
  topsep=0pt,
  parsep=0pt,
  partopsep=0pt
}

\title{Can Large Language Models Recover Semantic Optimization Opportunities That Compilers Miss?}
\author{
    Hailong Jiang\textsuperscript{\rm 1}\corresponding,
    Feng Yu\textsuperscript{\rm 1},
    Emran Hossain\textsuperscript{\rm 1},
    Jianfeng Zhu\textsuperscript{\rm 2},\\
    Mengfei Ren\textsuperscript{\rm 3},
    Qiang Guan\textsuperscript{\rm 2},
    Chunwei Xia\textsuperscript{\rm 4}
}

\affiliations{
    \textsuperscript{\rm 1}School of Computer Science, Information, and Engineering Technology, Youngstown State University\\
    \textsuperscript{\rm 2}Department of Computer Science, Kent State University\\
    \textsuperscript{\rm 3}Department of Computer Science, Baylor University\\
    \textsuperscript{\rm 4}School of Computer Science, University of Leeds\\
    Corresponding author: hjiang@ysu.edu
}

\begin{document}

\maketitle

\begin{abstract}
Optimizing compilers miss profitable transformations when their enabling semantics are absent from the analyzed program representation. 
We ask whether large language models (LLMs) can recover such semantics from heterogeneous C/C++ context and realize them as validated, contract-preserving artifacts. 
We introduce \segabench{}, an executable benchmark containing 100 synthetic and 20 source-backed cases spanning low-level assumptions, data-structure invariants, and high-level semantic lifting. Each case includes hidden enabling semantics, an oracle artifact, correctness and semantic validators, and a reproducible performance protocol. We evaluate five LLMs using five independent responses per case. 
The strongest model produces correct artifacts in 94.8\% of responses, achieves at least $1.05\times$ speedup in 83.3\%, and obtains a performance success on 93.3\% of cases. 
Nevertheless, correct artifacts often close only part of the oracle gap.
These results show that LLMs can complement compiler analysis as speculative semantic proposers, provided that their artifacts are validated and evaluated.
\end{abstract}


\section{Introduction}
\label{sec:intro}

\mypara{Motivation.}
C/C++ compilers optimize programs using the semantic information,
analyses, and transformations available in the compilation pipeline~\cite{burow2018taking,vafeiadis2015common,lossing2015data}.
Even aggressive compiler configurations such as \texttt{-O3} may leave
profitable transformations unexploited when their enabling semantics
cannot be established from the analyzed representation and scope~\cite{barany2018finding,d2015correctness}. 
Representative gaps include 
(i) missing context information. For example, by analyzing how data is used across multiple functions, an LLM may determine that two memory regions do not overlap and expose this information to the compiler, enabling vectorization~\cite{sheth2026preserving,anand2025iridescent};
(ii) data-structure invariants such as sortedness or diagonal structure; and 
(iii) program equivalences such as loops implementing scans or aggregations~\cite{gao2023systematic,blelloch1990prefix,steuwer2017lift}. 
Such evidence may be scattered across callers, tests, documentation, profiles, and configuration, requiring reasoning beyond an isolated function~\cite{paraskevopoulou2020verified,tang2026reasoning}.

Recent LLM-for-compilation work 
has focused on IR reasoning, low-level kernel generation, and iterative refinement with compilation and performance feedback~\cite{jiang2025can,chen2025cudallm,grubisic2024compiler}.
These works show that LLMs can reason over compiler artifacts and refine optimization candidates using compilation and performance feedback.
This motivates our hypothesis that LLMs may recover such missing optimization-enabling semantics from long program context and express them as high-level semantic artifacts.

\mypara{Research Questions.} For LLMs 
to serve as semantic bridges for compiler optimization, they must succeed in three steps: identify the semantic fact or better program equivalence that compilers cannot establish, realize it as a compiler-usable artifact, and translate that artifact into performance improvement.
We formulate these steps as three research questions.
\textbf{RQ1 (Semantic Identification)} asks whether 
LLMs can identify the semantic fact and its evidence;
\textbf{RQ2 (Artifact Realization)} asks whether 
LLMs can realize the fact as a valid artifact;
\textbf{RQ3 (Performance Realization)} asks whether valid artifacts 
deliver measurable speedups and how much of the oracle opportunity they recover.
Together, these three questions ask:
\begin{highquote}
Can LLMs recover semantic optimization opportunities missed by a strong C/C++ compiler baseline, and realize their enabling semantics as validated artifacts that deliver measurable speedups?
\end{highquote}
\vspace{-0.5em}

\mypara{Approach.}
To answer these RQs, we introduce \segabench{} (\textbf{se}mantic \textbf{ga}p \textbf{bench}mark),
an executable benchmark for evaluating semantic opportunity realization. Each case includes the original C/C++ program, relevant context, hidden target semantics and a reference implementation for evaluation, correctness checks, and a fixed performance-testing procedure. The oracle artifact serves as a validated reference: it passes all validators and achieves measurable speedup over the strongest compiled baseline.


\segabench{} contains two suites: a \textbf{Synthetic Suite} with 100 carefully designed cases covering 50 common optimization patterns, and a \textbf{Real-world Suite} with 20 cases collected from six widely used HPC projects. Together, they cover three semantic types: \textit{low-level assumptions}, \textit{data-structure invariants}, and \textit{high-level semantic lifting}.

\mypara{Evaluation Protocol.}
To evaluate LLMs, we use a blind, single-turn setup in which the model gives its answer without receiving any feedback from the evaluator. For each case, the model sees the original program and selected context, while the target semantics, reference solution, validation checks, and performance results are kept hidden.
The model can either decline to answer or identify a useful semantic fact or equivalent program behavior, explain the supporting evidence, and produce a source-level change. 
After the response is fixed, the evaluator turns the proposed change into a patch, compiles the modified program, checks its correctness, and measures the performance against the strongest compiled baseline. 
The model receives no feedback from the compiler, validators, reference solution, or profiler.


\mypara{Results.}
Across five LLMs and 3{,}000 responses, artifact realization is strongly model-dependent.
The strongest model produces correct, contract-preserving artifacts in 94.8\% of responses, achieves at least 1.05$\times$ speedup in 83.3\%, and, with five independent responses per case, finds a speedup on 93.3\% of the benchmark.
Yet correctness alone does not realize the full oracle opportunity: many correct artifacts close only part of the oracle gap, and performance success drops from 86.6\% on synthetic cases to 67.0\% on real-world cases.
These results highlight both the promise of LLM-guided semantic optimization and the need for validation and measurement.

\mypara{Contributions.}
This paper makes three contributions:
\begin{itemize}[label=\tiny$\blacksquare$]
    \item We formalize \emph{semantic optimization opportunities} missed by compiler baselines and formulate their recovery as an evidence-grounded LLM task. We organize their enabling semantics into three types and 50 recurring archetypes.

    \item We introduce \segabench{}, an executable benchmark comprising 100 synthetic and 20 source-backed HPC cases. 
    Each case provides hidden target semantics, an oracle artifact, functional and semantic validators, and fixed build and performance protocols.

    \item We design an evaluation protocol that separately measures semantic recovery, artifact correctness, and performance realization across models, suites, and semantic types.
\end{itemize}

\section{Task Definition}
\label{sec:problem}


\mypara{Semantic Optimization Opportunity.}
Let $P$ be a C/C++ program with a defined input domain and behavioral contract $\mathcal{D}$, and let $B$ denote the strongest evaluated compiler baseline for $P$. An \emph{enabling semantics} $S$ is a program property or equivalent computation that enables a more efficient implementation while producing the same observable output for every valid input, as illustrated in Figure~\ref{fig:benchmark-instance}.

We say that $(S^\star,A^\star)$ establishes a \emph{semantic optimization opportunity} for $P$ when three conditions hold. First, $S^\star$ is valid for $\mathcal{D}$, or its use is protected by a runtime guard with a correct fallback implementation. Second, the oracle artifact $A^\star$ realizes $S^\star$ while preserving the program contract:
\[
\forall x \in \mathcal{D}, \qquad
(P \oplus A^\star)(x) \simeq P(x),
\]
where $\simeq$ denotes observational equivalence under the benchmark contract. Third, the transformed program achieves a statistically significant speedup over the baseline:
\[
\operatorname{Speedup}(A^\star)
=
\frac{T(P,B)}
     {T(P \oplus A^\star,B)}
\ge p_{\mathrm{admit}},
\]
under the benchmark admission protocol.

Throughout this paper, a compiler \emph{misses} an opportunity when the strongest evaluated compilation of the original program leaves this admitted opportunity unexploited. 

\mypara{Semantic Artifacts.}
A \emph{semantic artifact} is a concrete source-level change that makes an optimization-enabling semantic fact or program equivalence explicit. Applying an artifact $A$ to a program $P$ yields $P \oplus A$. Artifacts may include qualifiers, assumptions, alignment annotations, guarded specializations, contracts, library substitutions, or rewrites. For example, non-aliasing may be expressed as \texttt{restrict} or a no-overlap guard, while an aggregation equivalence may be realized as an algorithmic rewrite.

Each benchmark case contains evaluator-only target semantics $S^\star$ and an oracle artifact $A^\star$, both hidden from the model. The oracle artifact is validated to preserve the benchmark contract and improve over the strongest baseline, witnessing a realizable opportunity rather than a global optimum. The model produces a candidate artifact $\widehat{A}$, which may differ from $A^\star$ but must realize a supported semantic claim and pass correctness and performance evaluation.

\mypara{LLM Task.}
For each case, the LLM receives the original program $P$ and an available context view $C$, while $S^\star$ and $A^\star$ remain hidden. The model may abstain or return a proposal
\[
    \widehat{Y}
    =
    (\widehat{S},\widehat{E},\widehat{A}),
\]
where $\widehat{S}$ is the proposed enabling semantics;
$\widehat{E}$ identifies supporting evidence and scope;
$\widehat{A}$ is the corresponding artifact.


\section{Taxonomy of Semantic Optimization Opportunity}
\label{sec:taxonomy}

\segabench{} organizes semantic optimization opportunities into three
types: low-level assumptions, data-structure invariants, and high-level semantic lifting. Within these types, we derive 50 archetypes by grouping recurring compiler limitations and performance transformations according to their enabling semantics, required evidence, and realizable artifact.
Each archetype therefore, represents a distinct semantic reason that an optimization remains unexploited and a corresponding way to realize it.
These archetypes guide the construction of the synthetic suite, while the same three types categorize the source-backed real-world cases. Table~\ref{tab:semantic_gap_taxonomy} summarizes the taxonomy.

\mypara{Low-Level Assumptions.}
Low-level assumptions refine a general C/C++ interface with a more
precise execution domain \cite{avans2025concepts,radtke2024extension}. They describe memory accesses, scalar values,
or iteration spaces, including non-aliasing pointers, aligned buffers,
bounded values, fixed strides, and fixed loop trip counts \cite{lattner2005macroscopic}. Supporting
evidence commonly appears in callers, allocators, validation logic, or
workload configuration. A property guaranteed by the program contract
can be realized as a scoped qualifier or assumption; a dynamically
established property requires a guard, a specialized path, and a correct
fallback. These artifacts can enable vectorization, unrolling, address
simplification, and check elimination \cite{1281665,zheng2025vectrans}.

\mypara{Data-Structure Invariants.}
Data-structure invariants describe semantic relationships among elements that give a concrete representation a stronger abstract meaning~\cite{liskov1987keynote}. Examples include
sorted or unique sequences, permutations, diagonal matrices, canonical
sparse layouts, injective index maps, and disjoint intervals \cite{di1989data}. Such
relations typically span the complete, data-dependent structure and are
not expressed by ordinary C/C++ element types or storage layouts \cite{kawaguchi2009type}. A
trusted constructor or complete invariant validator can establish the
stronger representation, after which a specialized algorithm operates
on it; a generic implementation handles inputs for which the invariant
does not hold \cite{gao2023spgemm}.

\mypara{High-Level Semantic Lifting.}
High-level semantic lifting recovers an equivalence between a low-level
implementation and a higher-level operation, algorithm, or fused
computation \cite{paraskevopoulou2020semantic}. Representative examples include recognizing loop nests as
group-by aggregation, prefix scan, sliding-window aggregation, stable
compaction, symmetric pairwise interaction, or a producer--consumer
pipeline \cite{blelloch1990prefix,steuwer2017lift,lattner2020mlir}. These equivalences span loops, calls, and intermediate data
structures, and their realization often requires whole-computation
restructuring \cite{bondhugula2008pluto}. Typical artifacts include equivalent source rewrites,
fused implementations, and specialized traversals. Semantic and
differential validation check equivalence with the original program,
including its ordering, boundary, and numerical behavior \cite{necula2000translation}.

\begin{table*}[t]
\centering
\caption{Taxonomy of semantic optimization opportunities. \tiny{(A ``miss'' is an opportunity left unexploited by the strongest compiler baseline.)}}
\label{tab:semantic_gap_taxonomy}
\footnotesize
\setlength{\tabcolsep}{4pt}
\renewcommand{\arraystretch}{1.}

\begin{tabularx}{\textwidth}{
    @{}
    >{\raggedright\arraybackslash}p{0.15\textwidth}
    >{\raggedright\arraybackslash}X
    >{\raggedright\arraybackslash}X
    >{\raggedright\arraybackslash}X
    @{}
}
\toprule
\textbf{Type}
& \textbf{Opportunity}
& \textbf{Why baseline misses it}
& \textbf{Artifact / optimization} \\
\midrule

Low-level assumptions (LLA)
& Memory, value, and iteration facts: non-aliasing, alignment, fixed strides/bounds, bounded indices.
& The interface permits more executions than the caller, allocator, or configuration actually allows.
& Assumptions or guarded specialization; enables vectorization, unrolling, address simplification, and check elimination. \\

\addlinespace[2pt]

Data-structure invariants (DSI)
& Global data relations: sortedness, uniqueness, diagonal structure, canonical sparsity, injective maps, disjoint intervals.
& Types and layouts expose storage, but not relations across elements.
& Invariant contract or validator with specialization; enables simpler access and specialized algorithms. \\

\addlinespace[2pt]

High-level semantic lifting (HSL)
& Program equivalences: scans, aggregation, sliding windows, compaction, pairwise interaction, producer--consumer fusion.
& Recognition spans loops, calls, and intermediates; realization requires restructuring the computation.
& Equivalent rewrite, fusion, or specialized traversal; reduces passes, intermediates, or asymptotic work. \\

\bottomrule
\end{tabularx}
\end{table*}

\section{Benchmark Design}
\label{sec:benchmark}

We turn the task in Section
~II into \segabench{}. 
It tests whether LLMs can recover optimization-enabling semantics missed by strong C/C++ compiler baselines and realize them as validated, contract-preserving artifacts. 
The benchmark includes a synthetic suite for controlled analysis and a real-world suite of source-backed application cases.

\begin{figure*}[t]
    \centering
    \includegraphics[width=0.94\textwidth]{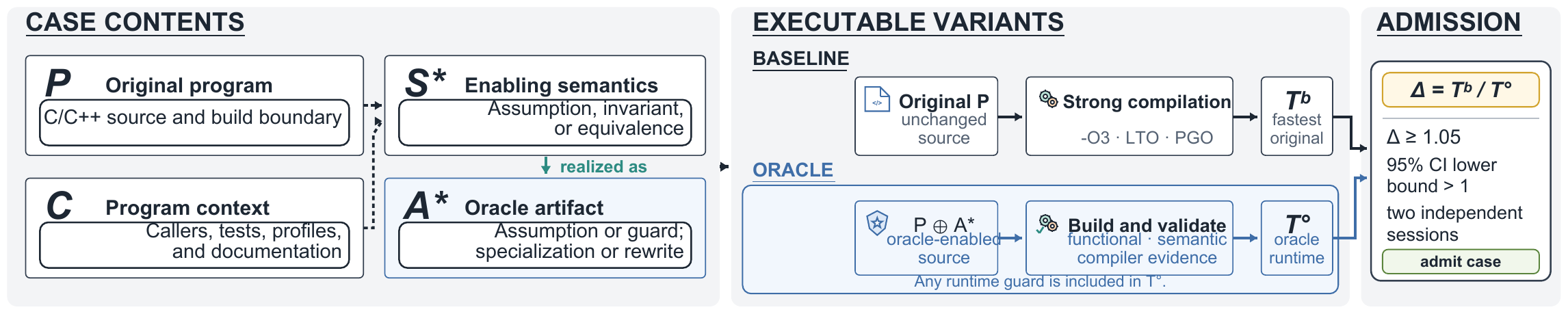}
    \caption{Structure and admission flow of a \segabench{} case.}
    \label{fig:benchmark-instance}
\end{figure*}

\mypara{Case Representation and Admission.}
As shown in Figure~\ref{fig:benchmark-instance}, each case is represented as
\[
    \mathcal{B}_i=(P_i,C_i,S_i^\star,A_i^\star,V_i,M_i),
\]
where $P_i$ is the original C/C++ program; $C_i$ is its context, such as callers, tests, documentation, profiles, or compiler reports; $S_i^\star$ is the hidden enabling property or program equivalence; $A_i^\star$ is the hidden oracle artifact; $V_i$ contains functional and property-specific validators; and $M_i$ fixes the build, workload, and timing protocol. 
Runtime guards and fallback paths, when needed, are part of the artifact.

A case is admitted only if the oracle artifact applies, compiles, preserves the program contract, passes all validators, and exposes the intended optimization. 
The strong baseline is the fastest original program compiled with \texttt{-O3}, LTO, or PGO~\cite{llvm_users_manual}. 
The oracle-realized program must achieve at least $1.05\times$ speedup, with the lower bound of its 95\% confidence interval above $1.0$, in two independent measurement sessions. 
Any guard or validation cost executed within the workload is included.

\mypara{Synthetic Suite.}
The synthetic suite contains 100 purpose-built cases organized into 50
semantic archetypes, with two distinct program instances per archetype.
Each case isolates one enabling property or equivalence while fixing its
program contract, context sources, oracle artifact, semantic validator,
and executable workload. The paired construction supports controlled
analysis across semantic categories, archetypes, and context views.

\mypara{Real-World Suite.}
The real-world suite contains 20 source-backed cases drawn from six
established HPC projects: 
HPCG, LAMMPS, LULESH, miniFE, RAJAPerf, and XSBench~\cite{heroux2013hpcg, thompson2022lammps,karlin2013lulesh,lin2015assessing,pearce2024rajaperf,tramm2014xsbench}. 
Each case is derived from a source-locked application hotspot and retains its relevant program context, build boundary, and workload. The cases represent 20 distinct archetypes and pass source and provenance verification, blinded human semantic review, oracle isolation, functional and semantic validation, compiler-evidence checks, and the common admission protocol.

The synthetic suite is distributed as 40 LLA, 30 DSI, and 30 HSL cases, while the real-world suite adds 20 source-backed cases, distributed as 8 LLA, 6 DSI, and 6 HSL cases.

\section{Evaluation}
\label{sec:evaluation}

We evaluate \segabench{} along the three RQs: semantic identification, artifact realization, and performance realization. This section describes the blinded protocol and the metrics used for each stage.

\mypara{Blinded Evaluation Protocol.}
For each case $i$, \segabench{} creates an anonymized request containing the original program $P_i$ and a selected context view $C_i^v$. 
The target semantics $S_i^\star$, oracle artifact $A_i^\star$, validators, and performance data are hidden from the model. 
The model may abstain or return one proposal consisting of a semantic claim, supporting evidence, and either a transformed \texttt{original.cpp} or an explicit \texttt{no\_artifact} response.

Each response is frozen before evaluation. 
The importer checks schema conformance and converts any returned source into an original-file-only patch. 
The evaluator applies the patch in an isolated workspace, compiles it, runs functional and property-specific semantic validators, and profiles every correctness-passing candidate. 
No compiler, validator, oracle, or profiler feedback is returned to the model.

\mypara{Models and Run Configurations.}
We evaluate {five LLM configurations: GPT-5.4 mini (gpt-5.4-mini-2026-03-17) \cite{openai2026gpt54mini}, GPT-5.6 Sol (gpt-5.6-sol)\cite{openai2026gpt56sol}, DeepSeek-V4 Pro (deepseek-ai/DeepSeek-V4-Pro)\cite{deepseek2026v4pro}, Llama 3.3 70B Instruct Turbo (meta-llama/Llama-3.3-70B-Instruct-Turbo)\cite{meta2024llama33}, and Ternary Bonsai 27B (Prism-ML/Ternary-Bonsai-27B) \cite{prismml2026ternarybonsai27b}. 
All models are evaluated using the same prompt, response schema, maximum output length, and frozen run plans.} 
For each model, we draw five
independent full-context responses for each of the 120 cases at
temperature $0.7$, yielding 600 requests per model and 3,000 requests
overall. Each request is stateless.
We use a maximum output length of 16,384 tokens. Responses are preserved and evaluated as returned, without repair, retry, or iterative optimization feedback.

All compilation, validation, and performance measurements are conducted on a Mac mini with a 10-core Apple M4 processor (four performance and six efficiency cores) and 16\,GB of memory, running macOS 26.5.2. We use Apple Clang 17.0.0 targeting ARM64. Benchmark cases and variants are evaluated serially. We do not apply explicit CPU-frequency locking or core-affinity controls.

\mypara{Baselines and Measurement.}
For case $i$, the baseline runtime $T_i^b$ is the minimum runtime of the original program compiled with \texttt{-O3}, LTO, and PGO. 
The oracle-enabled runtime $T_i^o$ represents the admitted semantic opportunity, and $T_{ir}^c$ is the runtime of the candidate produced by request $r$. 
All variants use the same compiler, workload, source boundary, and measurement configuration. 

Oracle admission uses two independent measurement sessions and a confidence-interval criterion, as described in Section~\ref{sec:benchmark}. Candidate results use the point estimate from the fixed candidate measurement protocol unless otherwise stated.

\mypara{End-to-End Success.}
We evaluate each proposal as a three-stage pipeline: 
\emph{semantic identification}, \emph{artifact realization}, and 
\emph{performance realization}. 
A proposal is an end-to-end (E2E) success only if it identifies the target 
semantics with supporting evidence, realizes it as a valid artifact, and 
achieves the fixed performance criterion over the strongest compiler 
baseline.

\mypara{RQ1: Semantic Identification.}
Two reviewers, blinded to model identity and candidate performance,
judge whether a proposal states $S_i^\star$ or an equivalent claim and
whether its cited evidence supports the stated scope. Disagreements are resolved through adjudication to obtain the final RQ1 labels. Schema conformance and citation bounds are checked mechanically. 

Let $s_r$ indicate that request $r$ passes this review. Over the set of
all requests $\mathcal{R}$, the primary RQ1 metric is
\[
\mathrm{RecoveryRate}
=
\frac{1}{|\mathcal{R}|}
\sum_{r\in\mathcal{R}} s_r.
\]

\mypara{RQ2: Artifact Realization.}
Reviewers first check that the artifact realizes the same claim assessed
in RQ1. The evaluator then checks that the patch applies, compiles, and
passes both functional and property-specific semantic validation. Let
$a_r$ indicate that these conditions hold.

We report the unconditional correct-artifact rate
\[
\mathrm{ArtifactRate}_{\mathrm{all}}
=
\frac{
\sum_{r\in\mathcal{R}} s_r a_r
}{
|\mathcal{R}|
},
\]
and the rate conditioned on successful semantic recovery,
\[
\mathrm{ArtifactRate}_{\mathrm{cond}}
=
\frac{
\sum_{r\in\mathcal{R}} s_r a_r
}{
\sum_{r\in\mathcal{R}} s_r
}.
\]
The first measures E2E artifact production, while the second
isolates artifact construction from semantic recovery.

\mypara{RQ3: Performance Realization.}
RQ3 evaluates whether semantically valid and correctness-preserving artifacts translate into runtime improvements. 
We first define
\[
\mathcal{C}
=
\{r\in\mathcal{R}:s_ra_r=1\},
\]
the set of requests that pass both RQ1 and RQ2.

Let $m_r=1$ when candidate timing completes and the quantities required
for performance classification are available. We define
\[
\mathcal{C}_{\mathrm{perf}}
=
\{r\in\mathcal{C}:m_r=1\}.
\]
For each $r\in\mathcal{C}_{\mathrm{perf}}$, we compute
\[
\mathrm{Speedup}_{ir}
=
\frac{T_i^b}{T_{ir}^c}
\]
and
\[
\mathrm{GapClosed}_{ir}
=
\frac{T_i^b-T_{ir}^c}
     {T_i^b-T_i^o}.
\]
Because the oracle is a validated reference rather than a global
optimum, $\mathrm{GapClosed}_{ir}>1$ is permitted.

Using $p=1.05$, we partition $\mathcal{C}_{\mathrm{perf}}$ into:
\begin{itemize}[label=\tiny$\blacksquare$]
    \item \textbf{Oracle-level realization:}
    $\mathrm{GapClosed}_{ir}\geq 1$;
    \item \textbf{Partial realization:}
    $\mathrm{Speedup}_{ir}\geq p$ \&
    $\mathrm{GapClosed}_{ir}<1$;
    \item \textbf{No meaningful realization:}
    $\mathrm{Speedup}_{ir}<p$.
\end{itemize}

These outcome rates use $|\mathcal{C}_{\mathrm{perf}}|$ as their
denominator and sum to one. We report
$|\mathcal{C}_{\mathrm{perf}}|/|\mathcal{C}|$ as performance-measurement
coverage. Candidates without a valid performance classification remain
failures in request-level E2E@$p$.



\mypara{Aggregation and Sampling.}
We report RQ1, RQ2, and E2E results overall and by benchmark suite and
semantic type. RQ1, RQ2, and E2E rates use all requests, whereas the
three RQ3 outcome shares use the performance-classifiable
correct-artifact set $\mathcal{C}_{\mathrm{perf}}$.

For repeated sampling, $\mathrm{Success}@k$ counts a case as successful
if at least one of its first $k$ responses passes RQ1 and RQ2 and
achieves $\mathrm{Speedup}\geq p$. We report $k=1,\ldots,5$ and use
$\mathrm{Success}@5$ as the summary case-level metric. Abstention,
schema compliance, artifact generation, compilation, functional
validation, and semantic validation are reported as diagnostic outcomes.


\section{Experimental Results}
\label{sec:findings}

Each of the five models produces five independent responses for each of the 120 cases, yielding 600 requests per model and 3,000 in total.
Recovery and $\mathrm{ArtifactRate}_{all}$ use all requests as the denominator, including abstentions and malformed responses; $\mathrm{ArtifactRate}_{cond}$ is conditioned on RQ1 recovery. RQ3 is evaluated on $\mathcal{C}=\{r:s_ra_r=1\}$. E2E@$1.05$ additionally requires at least $1.05\times$ speedup, while Success@$k$ counts a case when any of its first $k$ responses succeeds E2E.


Table~\ref{tab:overall_results} shows a wide capability gap. GPT-5.6 Sol is strongest at all three stages, with 95.0\% recovery, 94.8\% correct artifacts, and 83.3\% E2E performance success. DeepSeek-V4-Pro and GPT-5.4-mini reach 63.8\% and 41.7\% E2E@$1.05$, whereas Llama-3.3-70B-Turbo and Ternary-Bonsai-27B reach only 5.8\% and 5.0\%.
The benchmark therefore distinguishes not only semantic reasoning but also the ability to turn recovered semantics into executable artifacts.

\begin{table*}[!t]
\centering
\caption{Overall results over 600 requests per model.}
\label{tab:overall_results}
\footnotesize
\setlength{\tabcolsep}{4.5pt}
\renewcommand{\arraystretch}{1.06}
\begin{tabular}{@{}lrrrrr@{}}
\toprule
\textbf{Model} &
$\mathbf{RecoveryRate}$ &
$\mathbf{ArtifactRate}_{all}$ &
$\mathbf{ArtifactRate}_{cond}$ &
\textbf{E2E@$1.05$} &
\textbf{Success@$5$} \\
\midrule
GPT-5.4-mini          & \underline{86.0\%} & 54.8\% & 63.8\% & 41.7\% & 76.7\% \\
GPT-5.6 Sol               & \textbf{95.0\%} & \textbf{94.8\%} & \textbf{99.8\%} & \textbf{83.3\%} & \textbf{93.3\%} \\
DeepSeek-V4-Pro       & 84.0\% & \underline{81.3\%} & \underline{96.8\%} & \underline{63.8\%} & \underline{91.7\%} \\
Ternary-Bonsai-27B    & 14.8\% &  7.7\% & 51.7\% &  5.0\% & 20.8\% \\
Llama-3.3-70B-Turbo   & 21.0\% & 10.0\% & 47.6\% &  5.8\% & 12.5\% \\
\bottomrule
\end{tabular}
\end{table*}

Repeated sampling improves case coverage for every model. From $k=1$
to $k=5$, E2E coverage rises from 39.2\% to 76.7\% for
GPT-5.4-mini, from 82.5\% to 93.3\% for GPT-5.6 Sol, and from 67.5\% to
91.7\% for DeepSeek. At $k=5$, DeepSeek produces at least one
RQ1- and RQ2-passing artifact for 119 of 120 cases, but GPT-5.6 Sol retains
slightly higher performance coverage (112 versus 110 cases).

\begin{figure*}[!t]
    \centering
    \includegraphics[width=0.9\textwidth]{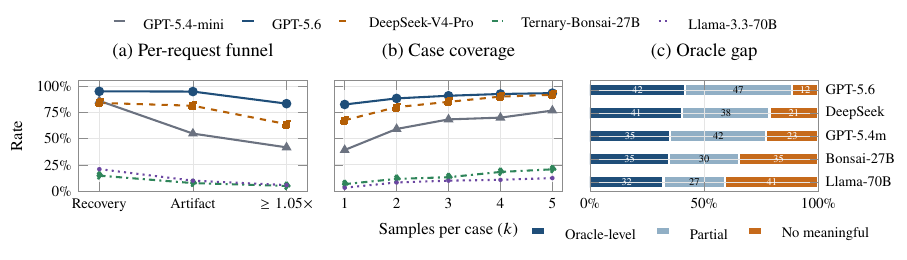}
\caption{Five-model results: (a) per-request RQ1, RQ2, and
E2E@$1.05$ rates; (b) case coverage versus samples per case; and
(c) RQ3 outcomes among performance-classifiable correct artifacts
$\mathcal C_{\mathrm{perf}}$.}
    \label{fig:overall_results}
    \label{fig:oracle_gap_results}
\end{figure*}

\mypara{RQ1: Semantic Opportunity Identification}
Blinded reviewers judge whether a response identifies the target
enabling semantics, or an equivalent claim, and grounds it in evidence
supporting its scope. GPT-5.6 recovers 570/600 opportunities (95.0\%),
followed by GPT-5.4-mini at 516/600 (86.0\%) and DeepSeek at 504/600
(84.0\%). Llama recovers 126/600 (21.0\%) and Ternary-Bonsai 89/600
(14.8\%). Their low rates primarily reflect missing reviewable claims:
Llama abstains in 474 responses, while Ternary-Bonsai has 511
non-recoveries, including 71 schema-invalid responses. Semantic
identification is thus reliable for the strongest models but is not a
general property of all evaluated LLMs.


\mypara{RQ2: Artifact Realization}
GPT-5.6 realizes 569/570 recovered opportunities as correct artifacts, giving $\mathrm{ArtifactRate}_{cond}=99.8\%$. DeepSeek realizes 488/504 (96.8\%), GPT-5.4-mini 329/516 (63.8\%),
Ternary-Bonsai 46/89 (51.7\%), and Llama 60/126 (47.6\%). Consequently, the correct-artifact rate ranges from 7.7\% to 94.8\%.

Execution failures further separate the models. All 570 GPT-5.6 artifacts compile and only one fails validation. Ternary-Bonsai has 20 compilation, six functional, and two semantic-validation failures;
Llama has 47 compilation, 18 functional, and one semantic-validation failure. Structured output compliance alone therefore does not imply artifact correctness.


\mypara{RQ3: Performance Realization}
The RQ3 correctness sets $\mathcal C$ contain 329 GPT-5.4-mini, 569 GPT-5.6, 488 DeepSeek, 46
Ternary-Bonsai, and 60 Llama responses. Of these, respectively, 76.0\%,
87.9\%, 78.5\%, 65.2\%, and 58.3\% achieve at least $1.05\times$.
Thus, even the two weaker models occasionally produce useful
artifacts, but their low RQ1 and RQ2 coverage reduces request-level
performance success to 5--6\%.

Figure~\ref{fig:oracle_gap_results} shows the fraction of the oracle
opportunity realized. GPT-5.6 has the smallest no-meaningful share
(11.7\%). This share increases to 21.4\% for DeepSeek, 22.8\% for
GPT-5.4-mini, 34.8\% for Ternary-Bonsai, and 40.7\% for Llama.
Median speedups among measured correct artifacts are $2.10\times$,
$2.02\times$, $1.77\times$, $1.63\times$, and $1.11\times$ for
GPT-5.6, DeepSeek, GPT-5.4-mini, Ternary-Bonsai, and Llama,
respectively. Conditional speedup must therefore be interpreted
together with E2E coverage.


\mypara{Synthetic versus real-world behavior.}
For the three strongest models, request-level E2E performance decreases
on the real-world suite. Ternary-Bonsai and Llama show higher
request-level rates on real-world cases because they return artifacts
more frequently. Nevertheless, median speedup among measured correct
artifacts decreases from synthetic to real-world cases for all five
models, from 1.95--7.29$\times$ to 1.04--1.29$\times$.
Suite comparison must therefore separate artifact coverage from the
speedup of correct artifacts.

\begin{figure*}[!t]
    \centering
    \includegraphics[width=0.9\textwidth]
        {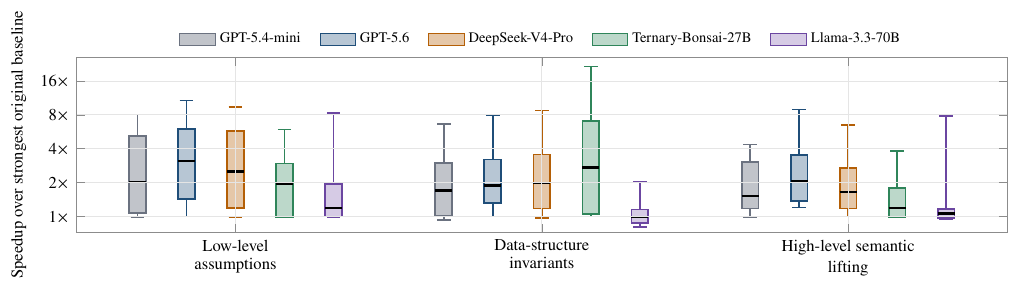}
\caption{Speedup of measured correct artifacts by semantic type and
model. Boxes show IQRs, center lines medians, and whiskers the
10th--90th percentiles; the dashed line marks baseline parity.}
    \label{fig:speedup_by_type}
\end{figure*}

\mypara{Speedup across Semantic Types.} Figure~\ref{fig:speedup_by_type} compares the runtime speedups of
correct, measured artifacts across the three semantic types and models.

Among the three strongest models, low-level assumptions produce the
largest median speedups: 2.03$\times$ for GPT-5.4-mini, 3.10$\times$
for GPT-5.6, and 2.51$\times$ for DeepSeek-V4-Pro. Ternary-Bonsai
instead peaks on data-structure invariants at 2.72$\times$, although
this estimate contains only 12 measured artifacts. Llama remains close
to baseline parity across all three types, with medians of
1.19$\times$, 0.99$\times$, and 1.07$\times$. Thus, performance
realization depends jointly on semantic type and model capability.

GPT-5.6 Sol achieves the highest median speedup on low-level assumptions and
high-level semantic lifting. DeepSeek-V4-Pro has the highest median on
data-structure invariants, although GPT-5.6 Sol has the higher geometric
mean, indicating a more right-skewed speedup distribution. Overall,
low-level semantic information produces the largest absolute runtime
benefits, whereas the performance of higher-level opportunities depends
more strongly on the model and individual benchmark case.

\begin{highquote}
\textbf{Answer to the central research question:\\}
Within the admitted opportunities in \segabench{}, capable LLMs can
recover and realize optimizations missed by compiler baselines, but the
capability is strongly model-dependent. GPT-5.6 Sol achieves 95.0\%
semantic recovery, 94.8\% correct artifacts, and 83.3\% E2E@$1.05$.
Performance is lower on source-backed cases for the strongest models,
and reliable use still requires validation and measurement.
\end{highquote}

\section{Discussion and Limitations}
\label{sec:discussion}

\mypara{Implications.}
Our results support a division of labor in which an LLM searches
heterogeneous program context for enabling semantics, validators establish
whether the proposed claim and artifact preserve the benchmark contract,
and the compiler performs downstream optimization and code generation.
This approach retains existing compiler infrastructure while extending
the semantic information available to it. However, the large differences
across models show that semantic opportunity realization is not yet a
uniform LLM capability. Repeated sampling improves case coverage, but it
also increases the need for reliable validation and candidate selection.
The lower speedups on real-world cases further indicate a transfer gap
between isolated opportunities and application-level performance.

\mypara{Limitations.}
\segabench{} currently covers C/C++, three semantic categories, and
source-backed cases from six HPC projects. Its real-world cases are
selected for reproducibility and measurable oracle speedup and may not
represent the full distribution of optimization opportunities in
production software. Performance is measured in one hardware and compiler
environment, so absolute speedups and model rankings may change across
platforms. Oracle artifacts are validated witnesses rather than global
performance upper bounds, and correctness guarantees apply only within
each case's defined contract and workload boundary. Finally, RQ1 relies
on blinded human judgments of semantic equivalence and evidence
sufficiency. Future work should extend the benchmark to additional
domains, languages, compilers, and hardware platforms, and investigate
machine-checkable contracts and automated candidate selection.

Candidate success uses point estimates, whereas oracle admission
additionally requires two independent sessions and a confidence-interval criterion; candidates near the $1.05\times$ threshold may therefore be sensitive to measurement noise.
\section{Related Work}
\label{sec:related}

\mypara{Learning-guided compiler optimization.}
Prior work applies machine learning to decisions within existing
compiler pipelines. MLGO integrates learned optimization policies into LLVM for decisions such as inlining, while CompilerGym provides an environment for learning-based exploration of compiler optimization sequences~\cite{trofin2021mlgo,
cummins2022compilergym}. Recent language-model-based approaches predict
optimization passes or optimized LLVM IR directly
~\cite{grubisic2024compiler, cummins2025llmcompiler}. HintPilot instead synthesizes compiler hints to steer existing compiler optimizations~\cite{jiang2026hintpilot}. These approaches operate over compiler-visible representations and available optimization mechanisms, either by improving optimization decisions or directly generating optimized code.
Our work instead studies whether LLMs can recover enabling semantics
that are not available to the original compilation in a usable form.

\mypara{LLMs for performance-oriented code generation.}
KernelBench evaluates LLM-generated GPU kernels using functional
correctness and speedup over PyTorch baselines, while CUDA-LLM
iteratively improves CUDA kernels using compilation, correctness, and
performance feedback~\cite{ouyang2025kernelbench,chen2025cudallm}.
These and related approaches similarly treat the LLM as a generator or tuner of a
replacement implementation. In contrast, \segabench{} begins with an
existing C/C++ program and asks the model to recover an optimization enabling semantic
property or program equivalence and realize it as a semantic artifact. The
existing compiler then remains responsible for downstream optimization and
code generation.

\mypara{Invariant discovery and verified transformation.}
Daikon dynamically infers likely program invariants from execution
traces~\cite{ernst2001dynamically}. Superoptimizers search or synthesize
equivalent program rewrites that improve a target cost metric 
~\cite{sasnauskas2017souper,liu2023minotaur}, while translation
validation systems such as Alive2 verify whether LLVM transformations
preserve program semantics~\cite{lopes2021alive2}. These techniques respectively
discover trace-supported properties, search a predefined transformation
space, or validate a supplied rewrite. \segabench{} connects these
concerns by evaluating evidence-grounded semantic recovery, executable
artifact construction, contract validation, and realized performance within
a common benchmark.
\section{Conclusion}
\label{sec:conclusion}

This paper asks whether LLMs can realize semantic optimization
opportunities missed by strong C/C++ compilers. We introduce
\segabench{}, comprising 100 synthetic and 20 source-backed cases with
hidden semantics, oracle artifacts, validators, and fixed performance
protocols. Across five LLMs, the strongest produces correct artifacts in
94.8\% of responses, achieves E2E@$1.05$ in 83.3\%, and succeeds on
93.3\% of cases with five samples. However, many artifacts close only
part of the oracle gap, and performance is lower on source-backed cases
for the strongest models. The large differences among models further
show that semantic opportunity realization is not yet a uniform LLM
capability. 
Overall, the results support a complementary workflow in which LLMs propose semantic artifacts, validators establish correctness, and compilers perform downstream optimization. Reliable use still requires contract-aware fallbacks and platform-specific performance measurement.



\bibliography{aaai2027}

@inproceedings{ouyang2025kernelbench,
  title     = {{KernelBench}: Can {LLM}s Write Efficient {GPU} Kernels?},
  author    = {Ouyang, Anne and
               Guo, Simon and
               Arora, Simran and
               Zhang, Alex L. and
               Hu, William and
               R{\'e}, Christopher and
               Mirhoseini, Azalia},
  booktitle = {Proceedings of the 42nd International Conference on Machine Learning},
  series    = {Proceedings of Machine Learning Research},
  volume    = {267},
  pages     = {47356--47415},
  publisher = {PMLR},
  year      = {2025},
  url       = {https://proceedings.mlr.press/v267/ouyang25a.html}
}

@article{chen2025cudallm,
  title   = {{CUDA-LLM}: {LLM}s Can Write Efficient {CUDA} Kernels},
  author  = {Chen, Wentao and
             Zhu, Jiace and
             Fan, Qi and
             Ma, Yehan and
             Zou, An},
  journal = {arXiv preprint arXiv:2506.09092},
  year    = {2025},
  doi     = {10.48550/arXiv.2506.09092},
  url     = {https://arxiv.org/abs/2506.09092}
}

@phdthesis{burow2018taking,
  title={Taking Back Control: Closing the Gap Between C/C++ and Machine Semantics},
  author={Burow, Nathan H},
  year={2018},
  school={Purdue University}
}

@inproceedings{vafeiadis2015common,
  title={Common compiler optimisations are invalid in the C11 memory model and what we can do about it},
  author={Vafeiadis, Viktor and Balabonski, Thibaut and Chakraborty, Soham and Morisset, Robin and Zappa Nardelli, Francesco},
  booktitle={Proceedings of the 42Nd Annual ACM SIGPLAN-SIGACT Symposium on Principles of Programming Languages},
  pages={209--220},
  year={2015}
}

@inproceedings{lossing2015data,
  title={From data to effects dependence graphs: source-to-source transformations for C},
  author={Lossing, Nelson and Guillou, Pierre and Amini, Mehdi and Irigoin, Fran{\c{c}}ois},
  booktitle={The 18th international workshop on compilers for parallel computing (CPC'15)},
  year={2015}
}

@inproceedings{d2015correctness,
  title={The correctness-security gap in compiler optimization},
  author={D'Silva, Vijay and Payer, Mathias and Song, Dawn},
  booktitle={2015 IEEE Security and Privacy Workshops},
  pages={73--87},
  year={2015},
  organization={IEEE}
}

@mastersthesis{sheth2026preserving,
  author = {Shravan Sheth},
  title = {Preserving Structural Alias Information Across the MLIR-to-LLVM Lowering Boundary},
  school = {California Polytechnic State University, San Luis Obispo},
  year = {2026},
  month = jun,
  type = {Master's thesis}
}

@article{anand2025iridescent,
  title={Iridescent: A framework enabling online system implementation specialization},
  author={Anand, Vaastav and Garg, Deepak and Kaufmann, Antoine},
  journal={arXiv preprint arXiv:2508.16690},
  year={2025}
}

@book{paraskevopoulou2020verified,
  title={Verified Optimizations for Functional Languages},
  author={Paraskevopoulou, Zoe},
  year={2020},
  publisher={Princeton University}
}

@article{tang2026reasoning,
  title={Reasoning compiler: LLM-guided optimizations for efficient model serving},
  author={Tang, Annabelle Sujun and Priebe, Christopher and Mahapatra, Rohan and Qin, Lianhui and Esmaeilzadeh, Hadi},
  journal={Advances in Neural Information Processing Systems},
  volume={38},
  pages={106904--106930},
  year={2026}
}

@article{radtke2024extension,
  title={An extension of C++ with memory-centric specifications for HPC to reduce memory footprints and streamline MPI development},
  author={Radtke, Pawel K and Barrera-Hinojosa, Cristian G and Ivkovic, Mladen and Weinzierl, Tobias},
  journal={arXiv preprint arXiv:2406.06095},
  year={2024}
}

@inproceedings{avans2025concepts,
  title={Concepts for designing modern C++ interfaces for MPI},
  author={Avans, C Nicole and Correa, Alfredo A and Ghosh, Sayan and Schimek, Matthias and Schuchart, Joseph and Skjellum, Anthony and Suggs, Evan D and Uhl, Tim Niklas},
  booktitle={European MPI Users' Group Meeting},
  pages={165--183},
  year={2025},
  organization={Springer}
}

@book{lattner2005macroscopic,
  title={Macroscopic data structure analysis and optimization},
  author={Lattner, Chris},
  year={2005},
  publisher={University of Illinois at Urbana-Champaign}
}

@article{zheng2025vectrans,
  title={Vectrans: Enhancing compiler auto-vectorization through llm-assisted code transformations},
  author={Zheng, Zhongchun and Wu, Kan and Cheng, Long and Li, Lu and Rocha, Rodrigo CO and Liu, Tianyi and Wei, Wei and Zeng, Jianjiang and Zhang, Xianwei and Gao, Yaoqing},
  journal={arXiv preprint arXiv:2503.19449},
  year={2025}
}

@inproceedings{liskov1987keynote,
  title={Keynote address-data abstraction and hierarchy},
  author={Liskov, Barbara},
  booktitle={Addendum to the proceedings on Object-oriented programming systems, languages and applications (Addendum)},
  pages={17--34},
  year={1987}
}

@article{di1989data,
  title={Data structures for compact sparse matrices representation},
  author={Di Felice, P and Agnifili, A and Clementini, Eliseo},
  journal={Advances in Engineering Software (1978)},
  volume={11},
  number={2},
  pages={75--83},
  year={1989},
  publisher={Elsevier}
}

@article{gao2023spgemm,
  author = {Gao, Jianhua and Ji, Weixing and Chang, Fangli and Han, Shiyu and Wei, Bingxin and Liu, Zeming and Wang, Yizhuo},
  title = {A Systematic Survey of General Sparse Matrix-Matrix Multiplication},
  journal = {ACM Computing Surveys},
  volume = {55},
  number = {12},
  articleno = {244},
  numpages = {36},
  year = {2023},
  month = mar,
  publisher = {Association for Computing Machinery},
  address = {New York, NY, USA},
  issn = {0360-0300},
  doi = {10.1145/3571157},
  url = {https://doi.org/10.1145/3571157}
}

@inproceedings{paraskevopoulou2020semantic,
  author = {Paraskevopoulou, Paraskevi and others},
  title = {Semantic Program Alignment for Equivalence Checking},
  booktitle = {Proceedings of the ACM SIGPLAN Conference on Programming Language Design and Implementation},
  year = {2020}
}

@inproceedings{bondhugula2008pluto,
  title={Pluto: A practical and fully automatic polyhedral program optimization system},
  author={Bondhugula, Uday and Hartono, Albert and Ramanujam, Jagannathan and Sadayappan, Ponnuswamy},
  booktitle={Proceedings of the ACM SIGPLAN 2008 Conference on Programming Language Design and Implementation (PLDI 08), Tucson, AZ (June 2008). Citeseer},
  volume={146},
  year={2008}
}

@manual{llvm_users_manual,
  title        = {Clang Users Manual},
  author       = {{LLVM Project}},
  organization = {LLVM Project},
  year         = {2025},
  url          = {https://clang.llvm.org/docs/UsersManual.html},
  note         = {Accessed July 27, 2026}
}

@techreport{heroux2013hpcg,
  title={HPCG benchmark technical specification},
  author={Heroux, Michael Allen and Dongarra, Jack and Luszczek, Piotr},
  year={2013},
  institution={Sandia National Laboratories (SNL-NM), Albuquerque, NM (United States)}
}

@article{thompson2022lammps,
  title={LAMMPS-a flexible simulation tool for particle-based materials modeling at the atomic, meso, and continuum scales},
  author={Thompson, Aidan P and Aktulga, H Metin and Berger, Richard and Bolintineanu, Dan S and Brown, W Michael and Crozier, Paul S and In't Veld, Pieter J and Kohlmeyer, Axel and Moore, Stan G and Nguyen, Trung Dac and others},
  journal={Computer physics communications},
  volume={271},
  pages={108171},
  year={2022},
  publisher={Elsevier}
}

@techreport{karlin2013lulesh,
  title={Lulesh 2.0 updates and changes},
  author={Karlin, Ian and Keasler, Jeff and Neely, J Robert},
  year={2013},
  institution={Lawrence Livermore National Laboratory (LLNL), Livermore, CA (United States)}
}

@article{lin2015assessing,
  title={Assessing a mini-application as a performance proxy for a finite element method engineering application},
  author={Lin, Paul T and Heroux, Michael A and Barrett, Richard F and Williams, Alan B},
  journal={Concurrency and Computation: Practice and Experience},
  volume={27},
  number={17},
  pages={5374--5389},
  year={2015},
  publisher={Wiley Online Library}
}

@inproceedings{pearce2024rajaperf,
  author    = {Olga Pearce and Jason Burmark and Rich Hornung and Befikir Bogale and Ian Lumsden and Michael McKinsey and Dewi Yokelson and David Boehme and Stephanie Brink and Michela Taufer and Tom Scogland},
  title     = {RAJA Performance Suite: Performance Portability Analysis with Caliper and Thicket},
  booktitle = {Proceedings of the 2024 IEEE/ACM International Workshop on Performance, Portability and Productivity in HPC (P3HPC)},
  year      = {2024},
  organization = {IEEE/ACM},
  note      = {Held in conjunction with the International Conference for High Performance Computing, Networking, Storage and Analysis (SC-W 2024)}
}

@article{tramm2014xsbench,
  title={XSBench-the development and verification of a performance abstraction for Monte Carlo reactor analysis},
  author={Tramm, John R and Siegel, Andrew R and Islam, Tanzima and Schulz, Martin},
  journal={The Role of Reactor Physics toward a Sustainable Future (PHYSOR)},
  year={2014}
}

@inproceedings{necula2000translation,
  title={Translation validation for an optimizing compiler},
  author={Necula, George C},
  booktitle={Proceedings of the ACM SIGPLAN 2000 conference on Programming language design and implementation},
  pages={83--94},
  year={2000}
}

@misc{openai2026gpt54mini,
  author       = {{OpenAI}},
  title        = {GPT-5.4 mini},
  year         = {2026},
  howpublished = {\url{https://developers.openai.com/api/docs/models/gpt-5.4-mini}},
  note         = {OpenAI API Documentation. Accessed July 2026}
}

@misc{openai2026gpt56sol,
  author       = {{OpenAI}},
  title        = {GPT-5.6 Sol},
  year         = {2026},
  howpublished = {\url{https://developers.openai.com/api/docs/models/gpt-5.6-sol}},
  note         = {OpenAI API Documentation. Accessed July 2026}
}

@misc{deepseek2026v4pro,
  author       = {{DeepSeek-AI}},
  title        = {DeepSeek-V4 Pro Model Card},
  year         = {2026},
  howpublished = {\url{https://huggingface.co/deepseek-ai/DeepSeek-V4-Pro}},
  note         = {Hugging Face Model Card. Accessed July 2026}
}

@misc{meta2024llama33,
  author       = {{Meta AI}},
  title        = {Llama 3.3 70B Instruct Model Card},
  year         = {2024},
  howpublished = {\url{https://huggingface.co/meta-llama/Llama-3.3-70B-Instruct}},
  note         = {Hugging Face Model Card. Accessed July 2026}
}

@misc{prismml2026ternarybonsai27b,
  author = {{Prism ML}},
  title  = {Ternary Bonsai 27B},
  year   = {2026},
  url    = {https://huggingface.co/prism-ml/Ternary-Bonsai-27B-gguf},
  note   = {Hugging Face Model Card. Accessed July 2026}
}

@article{trofin2021mlgo,
  title={Mlgo: a machine learning guided compiler optimizations framework},
  author={Trofin, Mircea and Qian, Yundi and Brevdo, Eugene and Lin, Zinan and Choromanski, Krzysztof and Li, David},
  journal={arXiv preprint arXiv:2101.04808},
  year={2021}
}

@inproceedings{cummins2022compilergym,
  title={Compilergym: Robust, performant compiler optimization environments for ai research},
  author={Cummins, Chris and Wasti, Bram and Guo, Jiadong and Cui, Brandon and Ansel, Jason and Gomez, Sahir and Jain, Somya and Liu, Jia and Teytaud, Olivier and Steiner, Benoit and others},
  booktitle={2022 IEEE/ACM International Symposium on Code Generation and Optimization (CGO)},
  pages={92--105},
  year={2022},
  organization={IEEE}
}

@article{grubisic2024compiler,
  title={Compiler generated feedback for large language models},
  author={Grubisic, Dejan and Cummins, Chris and Seeker, Volker and Leather, Hugh},
  journal={arXiv preprint arXiv:2403.14714},
  year={2024}
}

@inproceedings{cummins2025llmcompiler,
  author    = {Cummins, Chris and Seeker, Volker and Grubisic, Dejan and
               Rozi{\`e}re, Baptiste and Gehring, Jonas and
               Synnaeve, Gabriel and Leather, Hugh},
  title     = {LLM Compiler: Foundation Language Models for Compiler Optimization},
  booktitle = {Proceedings of the 34th ACM SIGPLAN International Conference
               on Compiler Construction},
  year      = {2025},
  pages     = {141--153}
}

@article{gao2023systematic,
  title={A systematic survey of general sparse matrix-matrix multiplication},
  author={Gao, Jianhua and Ji, Weixing and Chang, Fangli and Han, Shiyu and Wei, Bingxin and Liu, Zeming and Wang, Yizhuo},
  journal={ACM Computing Surveys},
  volume={55},
  number={12},
  pages={1--36},
  year={2023},
  publisher={ACM New York, NY}
}

@techreport{blelloch1990prefix,
  author       = {Guy E. Blelloch},
  title        = {Prefix Sums and Their Applications},
  institution  = {School of Computer Science, Carnegie Mellon University},
  number       = {CMU-CS-90-190},
  year         = {1990},
  month        = nov
}

@inproceedings{steuwer2017lift,
  title={Lift: a functional data-parallel IR for high-performance GPU code generation},
  author={Steuwer, Michel and Remmelg, Toomas and Dubach, Christophe},
  booktitle={2017 IEEE/ACM International Symposium on Code Generation and Optimization (CGO)},
  pages={74--85},
  year={2017},
  organization={IEEE}
}

@article{jiang2025can,
  title={Can Large Language Models Understand Intermediate Representations in Compilers?},
  author={Jiang, Hailong and Zhu, Jianfeng and Wan, Yao and Fang, Bo and Zhang, Hongyu and Jin, Ruoming and Guan, Qiang},
  journal={arXiv preprint arXiv:2502.06854},
  year={2025}
}

@article{ernst2001dynamically,
  title={Dynamically discovering likely program invariants to support program evolution},
  author={Ernst, Michael D and Cockrell, Jake and Griswold, William G and Notkin, David},
  journal={IEEE transactions on software engineering},
  volume={27},
  number={2},
  pages={99--123},
  year={2001},
  publisher={IEEE}
}

@article{sasnauskas2017souper,
  title={Souper: A synthesizing superoptimizer},
  author={Sasnauskas, Raimondas and Chen, Yang and Collingbourne, Peter and Ketema, Jeroen and Lup, Gratian and Taneja, Jubi and Regehr, John},
  journal={arXiv preprint arXiv:1711.04422},
  year={2017}
}

@article{liu2023minotaur,
  title={Minotaur: A SIMD-oriented synthesizing superoptimizer},
  author={Liu, Zhengyang and Mada, Stefan and Regehr, John},
  journal={arXiv preprint arXiv:2306.00229},
  year={2023}
}

@inproceedings{lopes2021alive2,
  title={Alive2: bounded translation validation for LLVM},
  author={Lopes, Nuno P and Lee, Juneyoung and Hur, Chung-Kil and Liu, Zhengyang and Regehr, John},
  booktitle={Proceedings of the 42nd ACM SIGPLAN International Conference on Programming Language Design and Implementation},
  pages={65--79},
  year={2021}
}

@inproceedings{barany2018finding,
  title={Finding missed compiler optimizations by differential testing},
  author={Barany, Gerg{\"o}},
  booktitle={Proceedings of the 27th international conference on compiler construction},
  pages={82--92},
  year={2018}
}

@inproceedings{jiang2026hintpilot,
  title={HintPilot: LLM-based Compiler Hint Synthesis for Code Optimization},
  author={Jiang, Hanyun and Yao, Peisen and Li, Kaiyue and Lin, Tingting and Wang, Chengpeng and Ren, Kui},
  booktitle={Findings of the Association for Computational Linguistics: ACL 2026},
  pages={24970--24986},
  year={2026}
}

@INPROCEEDINGS{1281665,
  author={Lattner, C. and Adve, V.},
  booktitle={International Symposium on Code Generation and Optimization, 2004. CGO 2004.}, 
  title={LLVM: a compilation framework for lifelong program analysis \& transformation}, 
  year={2004},
  volume={},
  number={},
  pages={75-86},
  doi={10.1109/CGO.2004.1281665}}

@inproceedings{kawaguchi2009type,
  title={Type-based data structure verification},
  author={Kawaguchi, Ming and Rondon, Patrick and Jhala, Ranjit},
  booktitle={Proceedings of the 30th ACM SIGPLAN Conference on Programming Language Design and Implementation},
  pages={304--315},
  year={2009}
}

@article{lattner2020mlir,
  title={MLIR: A compiler infrastructure for the end of Moore's law},
  author={Lattner, Chris and Amini, Mehdi and Bondhugula, Uday and Cohen, Albert and Davis, Andy and Pienaar, Jacques and Riddle, River and Shpeisman, Tatiana and Vasilache, Nicolas and Zinenko, Oleksandr},
  journal={arXiv preprint arXiv:2002.11054},
  year={2020}
}


\end{document}